\documentstyle[sprocl,epsfig]{article}

\def\be{\begin{equation}}
\def\ee{\end{equation}}
\def\bea{\begin{eqnarray}}
\def\eea{\end{eqnarray}}

\def\fICM{$f_{ICM}$}
\def\Tx{$T_{x}$}

\begin{document}

\title{Galaxy Cluster Baryon Fractions, Cluster Surveys and Cosmology\\
{\it Plenary contribution to PASCOS99 meeting, December, 1999}}

\author{Joseph J. Mohr$^{\ddagger}$\footnote{Chandra Fellow}, 
Zoltan Haiman$^{\dagger}$\footnote{Hubble Fellow} \& Gilbert P. 
Holder$^{\ddagger}$}

\address{$^{\ddagger}$University of Chicago Department of Astronomy and Astrophysics\\
$^{\dagger}$Princeton University Observatory}


\maketitle\abstracts{The properties of nearby galaxy clusters 
limit the range of cosmological parameters consistent with our 
universe.  We describe the limits which arise from studies of the 
intracluster medium (ICM) mass fraction \fICM\ and consideration of the 
possible sources of systematic error: $\Omega_{M}<0.44h_{50}^{-1/2}$ 
at 95\% confidence.  We emphasize that independent of Type Ia 
supernovae (SNe Ia) observations, this cluster study, taken 
together with published cosmic microwave background (CMB) anisotropy 
studies, indicates a non-zero 
quintessence or dark energy component $\Omega_{Q}>0$.\\
We then discuss future galaxy cluster
surveys which will probe the abundance of galaxy clusters to 
intermediate and high redshift.  
We investigate the sensitivity of these surveys to the cosmological
density parameter $\Omega_{M}$ and the equation of state parameter 
$w$ of any quintessence component.  In particular, we show that
cluster survey constraints from a proposed large solid angle X-ray 
survey are comparable in precision and complementary in nature
to constraints expected from future CMB anisotropy and SNe Ia studies.}

\section{Overview}
Galaxy clusters contain a wealth of accessible cosmological 
information.  Their masses range from 
$10^{14}$-$10^{15}$~$M_{\odot}$, 
making clusters the most massive collapsed or virialized 
objects in the universe.  Although numerous careful and varied studies 
of nearby clusters indicate that they are still accreting mass at the 
present epoch \cite{forman81,geller82,dressler88,mohr95}, it is 
evident that clusters exhibit striking 
regularity in scaling relations between mass, size and temperature 
\cite{mohr97,mohr99,horner99}.  In fact the scatter of clusters about 
the X-ray size-temperature relation is 15\%, comparable to the scatter 
of elliptical galaxies around the fundamental plane \cite{jorgensen95}.

Both these characteristics (accretion at the present epoch and tight 
scaling relations) are also exhibited by clusters formed within 
hydrodynamical simulations \cite{evrard93,evrard96,bryan98}, 
providing some confidence that the process of cluster formation is 
dominated by gravity and gas dynamics, and therefore simple enough
to to be effectively modeled in numerical simulations \cite{frenk99}.  
Ongoing observations with new X-ray observatories and proposed radio 
observatories will enable us to further improve our understanding of 
cluster formation and evolution \cite{roettiger98,roettiger99}.
Because galaxy clusters are sufficiently regular that their masses 
can be estimated from observables such as the ICM temperature \Tx, 
yields from cluster surveys of the high redshift universe are more 
readily interpreted and can, in principle, be used to constrain 
cosmological parameters. 

\section{Constraints on $\Omega_{M}$ from ICM Mass Fractions}
\label{sec:massfrac}

Galaxy clusters can be used to study the 
mix of baryonic and dark matter on scales of roughly 10~Mpc.   
Because there are no candidate mechanisms available to segregate 
baryons and dark matter on these scales \cite{evrard97}, it is often argued that the 
the baryon fraction within clusters $f_{cl}$ should reflect 
the universal baryon 
fraction $f_{B}\equiv\Omega_{B}/\Omega_{M}$, where $\Omega_{B}$ 
($\Omega_{M}$) is the 
cosmological density parameter of baryons (all clustered matter).  
Therefore, a measure of 
the cluster baryon fraction $f_{cl}$ can be combined with primordial 
nucleosynthesis constraints on the baryon to photon ratio and 
measurements of the CMB temperature to yield an estimate of the 
cosmological density parameter $\Omega_{M}=\Omega_{B}/f_{cl}$ 
\cite{white93}.

There are at least three reservoirs of baryons in 
galaxy clusters:  (1) the ICM, (2) the galaxies, and (3) dark baryons. 
The X-ray bremsstrahlung and 
recombination radiation from the ICM provides a precision tool for 
estimating the baryonic mass in the ICM reservoir.  The optical light 
emitted by stars within galaxies provides a more blunt estimate of the 
baryonic mass in galaxies (requires accurate estimates of the baryonic 
mass to light ratio in typical cluster galaxies), and to date there is 
no clean way of separating a possible dark baryonic component from the 
dominant dark matter reservoir within clusters.  Detailed studies of 
individual clusters tend to indicate that the ICM mass is several 
times larger than the galaxy mass \cite{david95}.Below
we describe a study which uses observations of the ICM baryon reservoir
to place an upper limit on $\Omega_{M}$.

\subsection{Study of the ICM in an X-ray Flux Limited Sample of 45 Clusters}

The ICM density profile $\rho(r)$ can be extracted from an image of the 
cluster X-ray emission, given an estimate of the mean ICM 
temperature \Tx, an emission model and some assumption about the 
cluster geometry.  The central 
X-ray surface brightness can be expressed 
\begin{equation}
I_{x}={1\over 2\pi(1+z)^{4}}\int_{0}^{\infty}\,dl\,{\rho(r)\over 
\mu_{e}\mu_{H}m_{p}^{2}} \Lambda(T_{x}),
\label{eq:Xray}
\end{equation}
where $\Lambda$ is an emission coefficient, $z$ is redshift, $m_{p}$ 
is the proton rest mass, $n_{i}\equiv\rho/\mu_{i}$, and $n_{i}$ is 
the number density of species $i$.  We use ROSAT PSPC observations of 
45 clusters from an X-ray flux limited sample with available data 
\cite{edge90}.
Within our chosen PSPC band (0.5:2~keV), the emission 
coefficient $\Lambda$ is extremely insensitive to ICM temperature \cite{mohr99}, 
and so one can accurately 
infer $\rho(r)$ without direct knowledge of the ICM temperature 
structure.  One does need to assume a geometry; we 
assume the clusters are spherical.  This introduces errors in the 
density inversion, and we characterize these errors by testing our 
analysis on mock observations of simulated galaxy clusters.  Using 
mock observations of 48 hydrodynamical simulations, we calculate that 
the cluster ICM mass within a radius $r_{500}$, roughly 
half the virial radius, can be estimated with an accuracy of 10\% \cite{mohr99}.

\begin{figure}[htb]
\vskip-0.3in
\hbox to \textwidth{\vbox to 2.5in{\hsize=2.4in
\psfig{figure=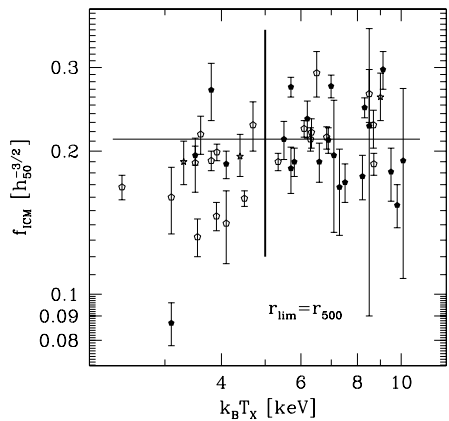,height=2.5in}\vfil}
\vbox to 2.5in{\hsize=2.25in\vskip5pt
\caption{Measured ICM mass fractions \fICM\ versus mean ICM temperature \Tx\  for
an X-ray flux limited sample of clusters.  
The mean \fICM\ for the 
clusters with $k_{B}T_{X}>5$~keV (vertical line) 
is 0.212$h_{50}^{-3/2}$ (horizontal line).  This measurement provides a 
lower limit on the fraction of matter within cluster virial regions 
which is baryonic.  Together with current best estimates of the scale 
of systematic errors on this upper limit, this measure of 
$\left<f_{ICM}\right>$ provides a 95\% 
confidence upper limit on the cosmological density parameter
of clustered matter:
$\Omega_{M}<0.44h_{50}^{-1/2}$.
\label{fig:ficm}}\vfil}\hfil}
\vskip-0.25in
\end{figure}

To estimate the ICM mass fraction \fICM\ we not only need the ICM 
density profile $\rho(r)$, but we also need the cluster binding mass.
We estimate $M_{500}$, the binding mass within $r_{500}$, by assuming 
the ICM is in hydrostatic equilibrium and is isothermal.  Departures 
from equilibrium and isothermality will introduce errors.  
A common temperature profile in clusters could 
potentially lead to systematic errors in our binding mass estimates.  
These are considered below in $\S$~\ref{sec:systematics}.

Fig.~\ref{fig:ficm} contains a plot of our \fICM\ measurements versus 
emission weighted mean ICM temperature \Tx\ in 45 clusters.  There are 
two important characteristics of the distribution of \fICM.  First, 
there is a weak, but statistically significant, tendency for low 
mass (low $k_{B}T_{x}$) clusters to have lower \fICM.  
The physics responsible for depleting the ICM in low mass clusters is 
thought to be preheating of the intergalactic medium by star formation 
within galaxies before the gas collapsed into the forming potential 
wells of clusters \cite{cavaliere98,ponman99}.  
Generally speaking, preheating of the gas prior to 
cluster formation or energy injection after cluster formation will 
have a larger effect on low mass clusters than on high.  Thus, in 
using \fICM\ to constrain the density of clustered matter, we restrict 
ourselves to the highest mass, hottest systems: $k_{B}T_{x}>5$~keV.

Second, splitting the 
sample at 5~keV (vertical line), we find the 
mean $f_{ICM}=(0.212\pm0.006)h^{-3/2}_{50}$ 
(the horizontal line).  This number is in reasonably good 
agreement with other estimates of \fICM 
\cite{white95,david95,arnaud99}.
Constraints on the primordial deuterium abundance from high redshift 
absorption systems provide an estimate of the baryon density 
parameter: $\Omega_{B}=0.076h_{50}^{2}$ \cite{burles98}.  Together 
with our lower limit on the cluster baryon fraction $f_{cl}$, this 
number implies an upper limit $\Omega_{M}<(0.36\pm0.01)h_{50}^{-1/2}$, 
where the uncertainty is only statistical.  
We require a factor of three systematic error 
to reconcile this number with $\Omega_{M}=1$ cosmological models.

\subsection{Discussion of Systematics}
\label{sec:systematics}
It is particularly interesting to consider systematics which could 
potentially make $\Omega_{M}=1$ models more consistent with our data.
There are several possible systematics:  
(1) any mechanism which enhances the cluster 
baryon fraction relative to the universal baryon fraction, 
(2) overestimating the ICM mass, and 
(3) underestimating the cluster binding mass $M_{500}$.

First, there are no known mechanisms for enhancing the baryon fraction by a 
factor of three on scales of 10~Mpc \cite{evrard97}.  
Second, plausible effects which 
would cause us to overestimate the ICM mass are (a) the effective area 
of the ROSAT PSPC is known to only 15\%, corresponding to a 7.5\% 
systematic uncertainty in our measured ICM masses, and (b) clumping 
or multiphase structure in the ICM would enhance the X-ray emission 
relative to single phase gas.  In fact, our hydro 
simulations indicate that the X-ray emission is enhanced by roughly 
20\% on average by clumping in the gas \cite{mathiesen99};  this leads 
us to overestimate the ICM mass by 10\%, on average (we have already 
corrected for this).  Interestingly, 
preliminary results from an ongoing analysis of an independent set of 
hydro simulations indicates that clumping from infalling substructure 
is present at about the same level as seen in our simulations 
\cite{bryan00}.  In addition, this clumping can potentially be 
addressed by comparison of ICM mass fractions derived from X-ray data 
and those derived from Sunyaev-Zel'dovich Effect (SZE) observations.  
Because the SZE is sensitive to a line integral of the $\rho T_{e}$, 
where $T_{e}$ is the electron temperature, clumping would likely have 
a much stronger effect on the X-ray measures than on the SZE.  A 
comparison of \fICM\ derived from an analysis of SZE observations of 
18 clusters provides no indication for systematically different 
results within an accepted range of the Hubble parameter $H_{0}$ 
\cite{grego00}.

Third, there are many possible effects which could systematically 
affect our binding mass estimates $M_{500}$; these include 
bulk flow or turbulence, support from magnetic fields, and ICM temperature 
profiles.  Indeed, analyses of some individual clusters provides 
binding masses derived from galaxy dynamics which are significantly 
higher than those derived from hydrostatic equilibrium.  However, a 
systematic study by the CNOC collaboration of 14 intermediate redshift 
clusters finds the ratio of galaxy dynamical to isothermal hydrostatic 
masses to be $1.04\pm0.07$ \cite{lewis99}.  Further studies using gravitational 
lensing and the spatially resolved ICM temperatures available from 
Chandra should provide much needed additional information.

Taking the 7.5\% systematic from the uncertainty in the PSPC 
effective area and the 7\% systematic uncertainty on the binding mass 
estimates, we estimate a total systematic uncertainty of 10\%.  This 
together with the observations outlined in the previous section leads 
to a 95\% confidence upper limit of $\Omega_{M}<0.44h^{-1/2}_{50}$.

\subsection{Two Independent Observational Arguments for $\Omega_{Q}>0$}

Studies of high redshift SNe Ia prefer cosmological models with 
$\Omega_{Q}>0$ \cite{schmidt98,perlmutter99}.  Both cluster baryon 
fraction arguments and mass to light studies \cite{carlberg97} favor 
low $\Omega_{M}$ models.  The mass to light ratio studies are more 
difficult to interpret, because the stellar populations of galaxies 
inside clusters differ significantly from those outside clusters.  
Nevertheless, these two approaches, subject to different systematics,
indicate $\Omega_{M}<<1$.  Together with 
constraints on CMB anisotropy\cite{dodelson00}, these clusters lead 
to the conclusion $\Omega_{Q}>0$, independent of the SNe Ia 
studies.  

\section{Cluster Surveys and Cosmology}

The relatively simple evolution (compared to galaxies) and 
regularity of galaxy clusters 
make them candidate tracer particles to use in measuring the 
volume--redshift relation.  This classical cosmological test \cite{tolman34} 
has been applied to galaxies with limited success, due at least in 
part to the complex relation between galaxy brightness and mass and 
the poorly understood evolution of the galaxy abundance 
\cite{loh86}(but see Newman \& Davis article for discussion of new 
approach being considered in the DEEP survey).  
Clusters are more amenable to these studies, because 
our theoretical understanding of their structure and evolution is more 
complete.  We don't expect the abundance of clusters to remain 
constant with redshift, but we can calculate its evolution, 
enabling the volume--redshift relation test.  
Moreover, the cosmological sensitivity of 
the abundance evolution itself provides additional leverage.

Galaxy cluster surveys of the nearby universe are an old endeavor \cite{abell58}; 
however, new technology and techniques are now making it 
possible to carry out extensive cluster studies of the intermediate and 
high redshift universe.  Studies of particular interest include: (1) 
a ~400~deg$^{2}$ serendipitous XMM cluster survey extending to
high redshift \cite{romer00}, (2) the 10$^{4}$~deg$^{2}$ survey of the 
nearby and intermediate redshift universe with the Sloan Digital 
Sky Survey (SDSS) \cite{annis00}, 
(3) a ~12~deg$^{2}$ interferometric SZE survey of the high 
redshift universe \cite{mohr99,holder00}, and 
(4) a 10$^{4}$~deg$^{2}$, deep X-ray survey of the nearby and intermediate 
redshift universe.

\begin{figure}[htb]
\vskip-0.2in
\hbox to \textwidth{\vbox to 2.5in{\hsize=2.4in
\psfig{figure=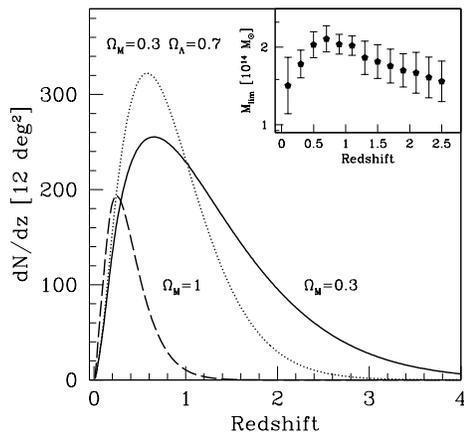,height=2.5in}\vfil}
\vbox to 2.5in{\hsize=2.25in
\caption{We plot the expected redshift distribution of a proposed SZE
survey for three different cosmological models.  The inset (upper 
right) shows the mass of clusters detectable at 5$\sigma$ 
significance as a function of redshift.  Note the mild redshift 
sensitivity of this mass threshold; this is the unique characteristic
of an SZE 
survey.  This particular proposed interferometric SZE survey will 
cover 12~deg$^{2}$ in one year, and it should yield a sample of 
roughly 400 clusters if the currently favored cosmological model 
($\Omega_{M}=0.3$, $\Omega_{\Lambda}=0.7$, $\sigma_{8}=1$) is correct.
\label{fig:SZEsurvey}}\vfil}\hfil}
\vskip-0.20in
\end{figure}

Fig~\ref{fig:SZEsurvey} contains expectations for the redshift 
distribution of the interferometric SZE survey proposed by J. Carlstrom 
and collaborators.  Because the SZE is a distortion of the CMB spectrum 
caused by inverse Compton scattering of CMB photons with hot 
electrons in the ICM, the SZE doesn't suffer from the cosmological 
dimming that any light source experiences.  This fact makes the SZE 
ideally suited to studies of massive structures in the high redshift 
universe.  The inset of Fig~\ref{fig:SZEsurvey} contains the mass of a 
cluster which would be detected with 5$\sigma$ significance as a 
function of redshift\cite{holder00}.  Interestingly, this 
limiting mass is relatively 
constant with redshift, and it corresponds to a very low mass galaxy 
cluster;  thus, the proposed Carlstrom survey will enable us to probe 
the universe for clusters to the very moment of their emergence.  
As described below, this fundamental observation will provide a 
powerful test of structure formation models and allow precision 
measurements of several cosmological parameters.

\begin{figure}[htb]
\vskip-0.2in
\hbox to \textwidth{\hskip-0.25in\vbox to 4in{\hsize=2.3in
\psfig{figure=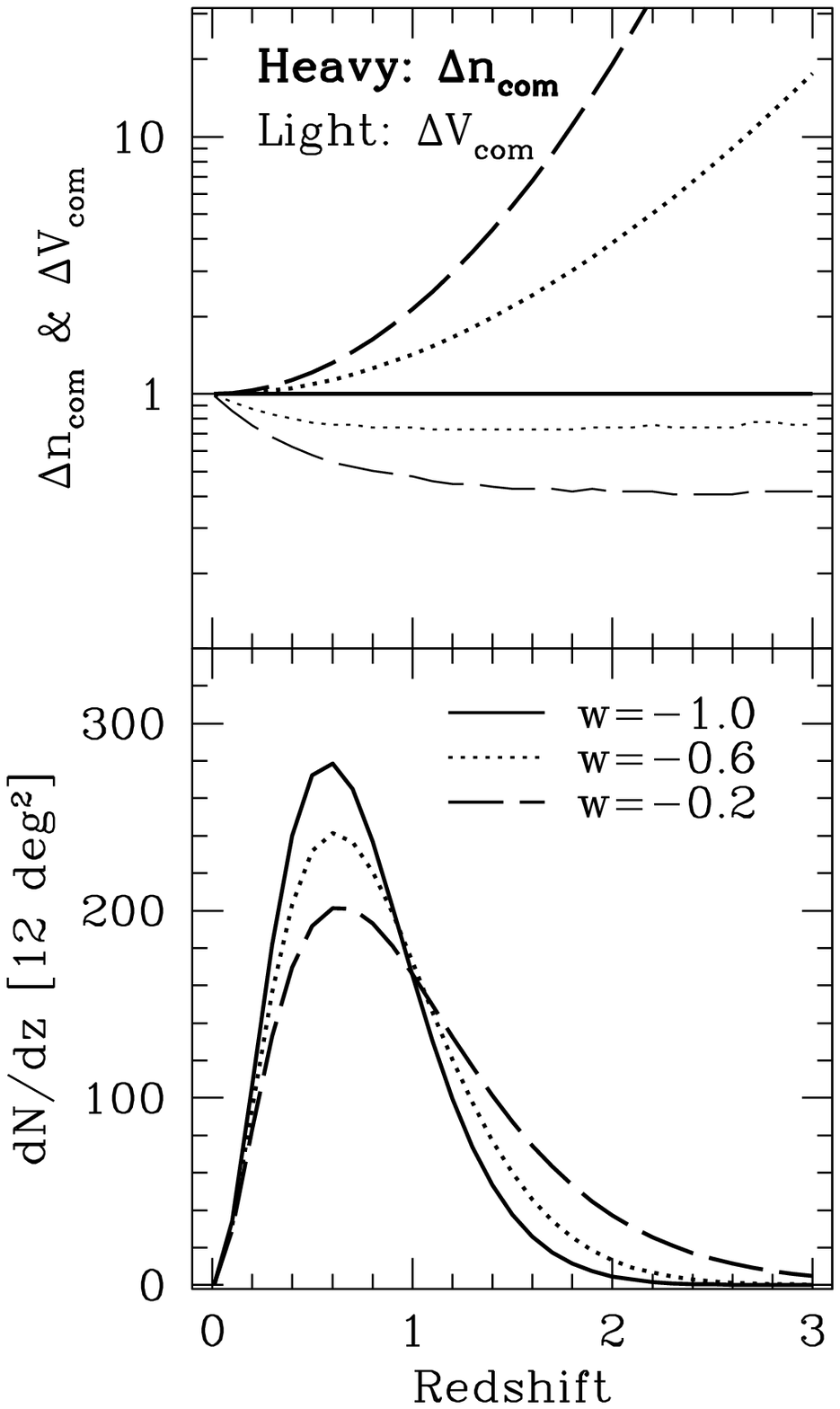,height=4.4in}\vfil}\hskip+0.05in
\vbox to 4in{\hsize=2.3in
\psfig{figure=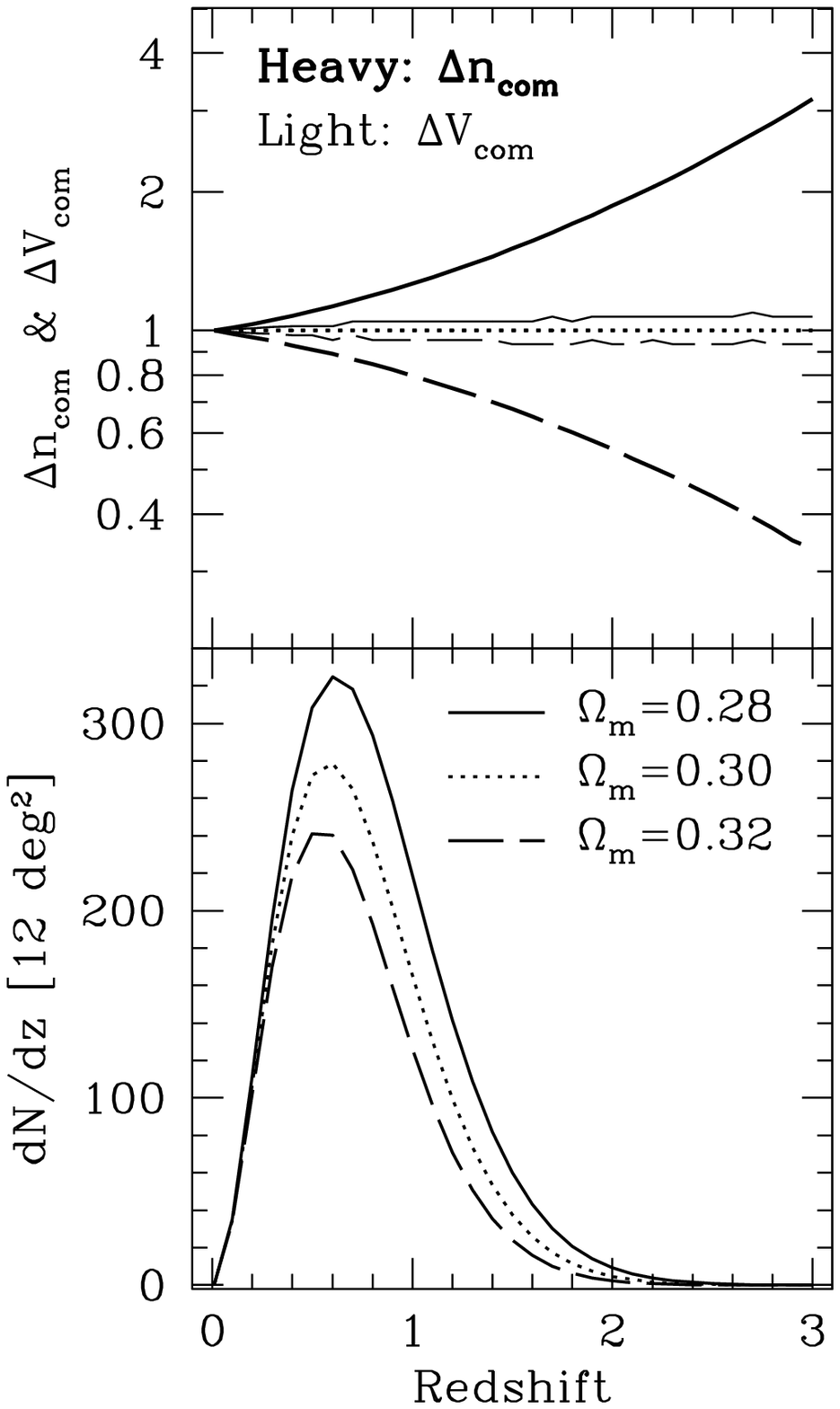,height=4.4in}\vfil}}
\vskip-0.0in
\caption{We show the sensitivity of an SZE survey to changing $w$ 
(left) and changing $\Omega_{M}$ (right).  These differences are 
separated (top) into the effects on the volume surveyed
and the abundance evolution.  Both the volume (light line) and 
abundance (heavy line) for each model are shown relative the the 
volume and abundance of the fiducial model ($w=-1$ and 
$\Omega_{M}=0.3$).  Note the clear differences between changes in $w$ 
and changes in $\Omega_{M}$, indicating that in principle $w$ and 
$\Omega_{M}$ can be determined simultaneously.
\label{fig:paramdepend}}
\vskip-0.20in
\end{figure}

\subsection{Constraining the Equation of State Parameter $w$}
Because there are now two independent observational arguments for 
a non-zero quintessence or dark energy component $\Omega_{Q}$, further 
work is required not only to test this conclusion, but also to 
make measurements of the equation of state parameter $w\equiv p/\rho$ of 
this component.  In principle the proposed SZE survey can do just 
this\cite{haiman00}, because the equation of state of the dark energy 
detemines how its energy density evolves $\rho_{Q}\propto 
R^{-3(1+w)}$, where $R$ is the scale factor.  This evolution affects the 
expansion history of the universe, which is coupled to the 
volume-redshift relation and the growth rate of density 
perturbations\cite{peebles80}.

Fig~\ref{fig:paramdepend} contains a plot of the $\Omega_{M}$ and $w$ 
dependence of the yields of the SZE survey in the case that the mass 
limit is taken to be a constant 
$M_{lim}=2\times10^{14}h_{50}^{-1}M_{\odot}$.  Note 
(lower left) that increasing $w$ from -1 (the $\Lambda$ case) to -0.2 
( with fixed $\Omega_{M}=0.3$) decreases the number of cluster 
expected at intermediate redshifts, 
but increases the number expected at high redshift.  This is explained 
in the upper left panel;  at low redshift, the 
larger surveyed volume of the $w=-1$ model is the dominant factor, whereas at 
higher redshift the evolution of the abundance plays a larger--- and 
eventually dominant--- role.  The right panel contains expected yields 
as a function of $\Omega_{M}$ (with fixed $w=-1$).  Lower $\Omega_{M}$ yields 
are larger, 
and they grow ever larger fractionally with redshift.  This is 
essentially the result of the evolution of abundances, because the volumes 
probed are very similar in these models (upper right).  The slower 
growth of density perturbations in low $\Omega_{M}$ models means that the 
abundance changes more slowly, and we expect to see clusters to 
higher redshift.
For more realistic tests which include the cosmological dependence of 
the limiting mass $M_{lim}(z)$, please see Haiman, Mohr \& Holder (2000).

\subsection{Complementary, High Precision Cosmological Constraints}

Finally, we turn to consider the cosmological constraints possible 
from the large solid angle, deep X-ray survey proposed by G. 
Ricker, D. Lamb and collaborators.  Although this survey extends only 
to a redshift $z\sim0.7$, it will detect 10$^{4}$ clusters 
and provide temperature measurements for $\sim$2,000 of these.  
These large numbers make for strikingly precise cosmological 
constraints, and having the \Tx\ and luminosity measurements for these 
clusters makes it possible to test for unusual evolution models while 
constraining cosmological parameters \cite{haiman00}.

\begin{figure*}[htb]
\vskip-0.2in
\hbox to \textwidth{\vbox to 2.5in{\hsize=2.4in
\psfig{figure=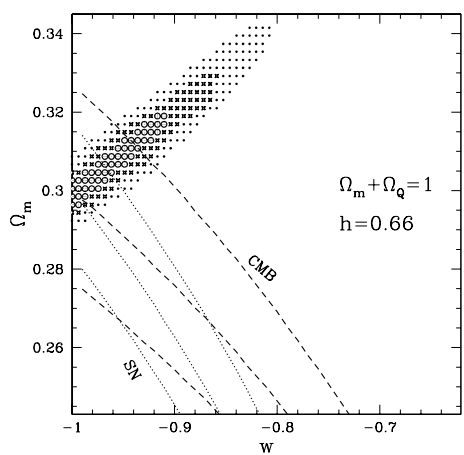,height=2.5in}\vfil}
\vbox to 2.5in{\hsize=2.25in\vskip-3pt
\caption{Here are estimates of the power contained in a 
large solid angle X-ray cluster survey 
to measure $\Omega_{M}$ and $w$ simultaneously.
The dots mark the 1$\sigma$, 2$\sigma$ and 3$\sigma$ confidence regions 
with which models can be differentiated from a fiducial model 
($\Omega_{M}=0.3$, $w=-1$, $\sigma_{8}=1$).  We show 
only the slice of parameter space with $h=0.66$; all models
have $\Omega_{M}+\Omega_{Q}=1$.  For comparison, the 
dashed lines indicate the constraint from a 1\% measurement 
of the location of the first doppler peak in the CMB anisotropy 
spectrum, and the dotted lines indicate the constraints from a 1\% measurement of 
the distance to redshift $z=1$.
In combination with either the CMB or SNe Ia studies, 
cluster surveys enhance the constraints on $w$.
\label{fig:XRzoom}}\vfil}\hfil}
\vskip-0.15in
\end{figure*}

Fig~\ref{fig:XRzoom} provides an estimate of the extent to which the 
clusters from this proposed survey will allow one to simultaneously 
constrain $w$ and $\Omega_{M}$.  In these models we take 
$\Omega_{M}+\Omega_{Q}=1$, and we require the local abundance of 
cluster above some mass threshold to be the same in each model.  For 
this figure we assume some fiducial model: $\Omega_{M}=0.3$, 
$\Omega_{Q}=0.7$, $\sigma_{8}=1$, $w=-1$ and $h=0.66$.  We then use the 
total number of expected clusters and their redshift distribution to 
quantify the differences between any given model and the fiducial 
model (redshifts for these clusters will be extracted from the SDSS 
imaging and spectroscopic survey).  Note the tight, simultaneous 
constraints on $w$ and 
$\Omega_{M}$.  For comparison, we show (dashed) the constraints 
corresponding to a 1\% measurement of the location of the first 
Doppler peak in the CMB anisotropy spectrum, and (dotted) a 1\% 
measurement of the distance to redshift $z=1$.  Note that the
cluster survey does better than a 1\% measurements to $z=1$.  
In addition, the parameter degeneracies from our cluster 
study and these future CMB and SNe Ia constraints are roughly orthogonal, 
making the cluster survey complementary to either of the other 
studies.  This orthogonality stems from the cluster survey 
sensitivity to abundance evolution.

\subsection{Systematic Effects and Non-standard Evolution}
Studies of the evolution of cluster structure are an important 
component of the effort to use cluster surveys to constrain 
cosmological parameters.  The structure of cluster virial regions 
can affect the relation between the cluster virial mass and observables 
like X-ray luminosity, emission weighted ICM temperature and SZE flux.
Very specific evolution models follow from 
theoretical work, and the observations 
required to test these models are now more readily available.
Nevertheless, still uncertain physics, such as the effects of heating of the 
intergalactic medium before cluster formation,
provides a potential source of systematic error in 
interpreting cluster surveys; we and others are currently studying 
these effects using hydrodynamical simulations\cite{bialek00}.  
The enormous successes in modeling galaxy clusters to 
date and the vast array of data soon to be available on intermediate and 
high redshift clusters lead me to adopt an optimistic view; I suspect that through 
further enhancing the hydrodynamical cluster simulations, we can tackle the 
finer points of cluster evolution and cosmology simultaneously.

\section*{Acknowledgments}
JJM is supported by the Chandra Fellowship grant PF8-1003, awarded 
through the Chandra Science Center.  The Chandra Science Center is 
operated by the Smithsonian Astrophysical Observatory for NASA under 
contract NAS8-39073.
ZH is supported by NASA through the Hubble Fellowship grant 
HF-01119.01-99A, awarded by the Space Telescope Science Institute, 
which is operated by the Association of Universities for Research in 
Astronomy, Inc., for NASA under constract NAS 5-26555.

\end{document}